\documentclass[12pt,preprint]{aastex}
\usepackage{graphicx}

\newcommand{\lp}{LP 944--20}
\newcommand{\pimms}{PIMMS}
\newcommand{\xmm}{\emph{XMM--Newton}}

\shorttitle{XMM-Newton observations of the nearby brown dwarf LP 944-20}
\shortauthors{Eduardo L. Mart\'\i n, Herv\'e Bouy}

\begin{document}

\title{XMM-Newton observations of the nearby brown dwarf LP 944-20}
\author{Eduardo L. Mart\'\i n}
\affil{Institute for Astronomy, 2680 Woodlawn Drive, Honolulu Hawai'i, 96822, USA}
\email{ege@teide.ifa.hawaii.edu}

\and
\author{Herv\'e Bouy}
\affil{E.S.O, Karl Schwarzschildstra\ss e 2, D-85748 Garching, Germany}
\email{hbouy@eso.org}



\begin{abstract}
The nearby ($d=5.0$ pc) brown dwarf \lp\ was observed with the \xmm\ satellite on 07 January 2001. 
The target was detected with the Optical Monitor ($V=$16.736$\pm$0.081), but it  
was not detected during the $\approx 48$ ks observation with the X-ray telescopes. 
We determine a $3\sigma$ upper limit for the X-ray emission from this object 
of $L_{X}<3.1 \times 10^{23}$ $ergs \cdot s^{-1}$, 
equivalent to a luminosity ratio upper limit of $\log{(L_{X}/L_{bol})} \le -6.28$. 
This measurement 
improves by a factor of 3 the previous \emph{Chandra} limit on the quiescent X-ray 
flux. This is the 
most sensitive limit ever obtained on the quiescent X-ray emission of a brown dwarf. 
Combining the  \xmm\ data with previous \emph{ROSAT} and \emph{Chandra} data, 
we derive flare duty cycles as a function of their luminosities. We find that very 
strong flares [Log$(L_X/L_{bol})>-$2.5] are very rare (less than 0.7\% of the time). 
Flares like the one detected by Chandra [Log$(L_X/L_{bol})=-$4.1] have a duty cycle 
of about 6\%, which is lower than the radio flare duty cycle ($\sim$13\%).  
When compared with other M dwarfs, \lp\ appears to be rather inactive in X-rays despite of its relative youth, fast rotation and its moderately strong activity at radio wavelengths. 

\end{abstract}


\section{Introduction}

\lp\ (=BRI 0337-3535) is an isolated nearby brown dwarf (Mass$<$0.075~M$_\odot$) first identified by \citet{luyten1975}. 
Spectroscopic observations reported by \citet{tinney98} give a 
spectral type of dM9, an age of about 500 Myr and a mass of about 0.065~M$_\odot$ 
using the lithium test proposed by \citet{magaz93}. 
Because of its substellar mass and its proximity to the Sun, 
this object is a benchmark in the study of very low-mass objects. 
\lp\ was observed in the X-Ray regime for the first time with the \emph{ROSAT} satellite by \citet{neuhauser99}, but was not detected. It has been detected with the \emph{Chandra} satellite \citep{rutledge2000} in december 1999 during an X-ray flare of duration $1\sim2$ hours. Outside the flare, \lp\ was not detected with a 3 $\sigma$ upper limit on the emission at $L_{X}<1.0 \times 10^{24}$ $ergs \cdot s^{-1}$. 
Both quiescent and flaring radio emission have been observed with the VLA \citep{berger2001} . 

X-ray emission is widespread among fully-convective M dwarfs (spectral type M3 and later), and it  
is frequently variable \citep{fleming1993, marino2000, feigelson2002}. X-ray observations of ultracool dwarfs (spectral type M7 and later) are still 
very scarce. Strong variability has been observed in a few objects \citep{rutledge2000, fleming2000, sch02}. 
It is thought that the X-ray photons are emitted from a hot corona. The properties of coronae (permanent or transient) 
in ultracool dwarfs are not well understood. 

\lp\ was targeted for observations with \xmm\ for two reasons mainly, first 
to try to catch it during a flare, which would have allowed 
to obtain spectroscopy if a flare had occurred, and second to search for quiescent X-ray emission. 
Unfortunately, we did not detect \lp\ at all. Nevertheless, 
we improve the value of the upper limit on the quiescent X-ray emission 
from \lp, and we derive X-ray flare rates as a function of X-ray luminosity that may be useful for planning future observations of this brown dwarf.

\section{Observations}

\xmm\ observed \lp\ on 2001 January 07-08, between 14:21:05 and 04:43:52 UTC for 51767 s. The pointing position was $R.A.=03 \mathrm{h} 39 \mathrm{min} 34.60 \mathrm{sec}$ and $Dec=-35\degr 25''51.0'$ Epoch 2000, according to the Simbad database.\footnote{Simbad database, operated at CDS, Strasbourg, France}

\subsection{Optical Data and Source Identification}

Besides its three X-ray telescopes, \xmm\ also has a 30-cm optical-UV telescope, providing the possibility to observe simultaneously in the X-ray and optical-UV regimes.
The Optical Monitor (\emph{OM}) was used with the $V$ filter for four exposures of 5000~s each in imaging mode. 
The OM detector has a format of $2048 \times 2048$ pixels, each $0.5'' \times 0.5''$. The field of view is therefore $17' \times 17'$.

We used the data given by the OM to confirm the presence of the source in the field of view. 
The coordinates we used to identify \lp\ are different from the pointing coordinates. We used the more recent coordinates given 
by the astrometry of \citet{rutledge2000} after their observation of \lp\ with \emph{Chandra} during a X-ray flare in 1999 December. 
These coordinates are: $R.A.=03\mathrm{h}39\mathrm{min}35.16\mathrm{sec} \pm 0.1''$ and $Dec=-35^{\circ}25''44.0' \pm 0.1''$,  
epoch 1999.95.

\begin{figure}
\figurenum{1}
\label{fig1}
\plotone{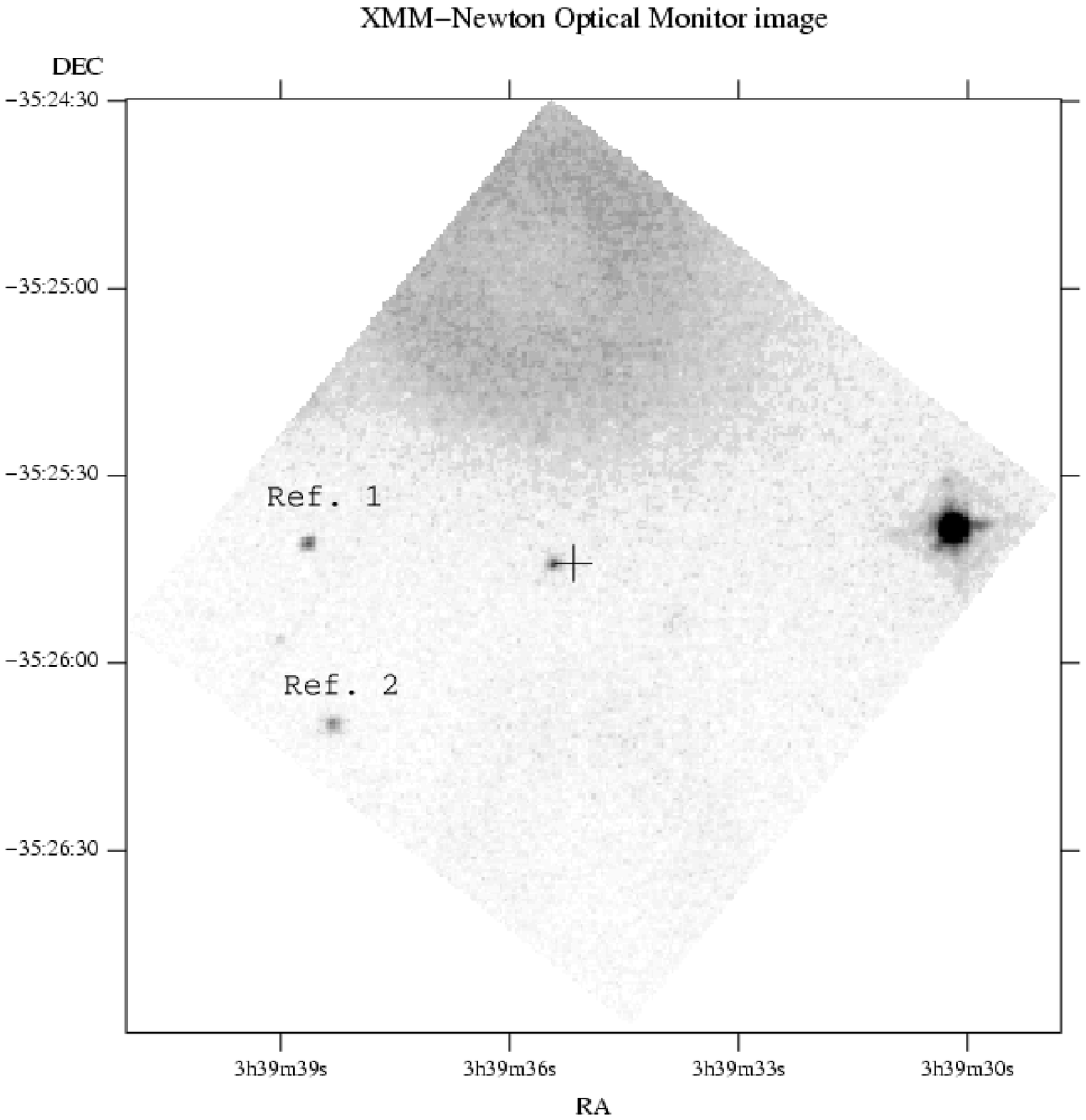}
\vspace*{-2in}
\caption{\small Indentification of \lp\ in the field of view of the Optical Monitor Data. The plus indicate the expected coodinates of \lp. 
The diffuse bright area on the top of the image is an artifact.}
\end{figure}

We identified \lp\ in the field of view by plotting these coordinates on the four images given by the OM. 
The result is shown in Figure \ref{fig1}. We found a point-source at a distance of $0.4''\pm 0.1''$ from the expected 
coordinates of \lp. This offset is small in comparison with the $<2''$ known systematic uncertainties in the \xmm\ preprocessing analysis astrometry \citep{xmmuhb}. Moreover a small offset was also  expected because of the proper motion of \lp\ since it was observed by \citet{rutledge2000} more than one year ago. 
The offset in the optical position is consistent with the known proper motion of \lp. 
We were also able to identify the three other sources in the field of view simply by direct comparison with the image obtained by \citet{rutledge2000}.
We therefore identify this optical source with the brown dwarf \lp\ on the basis of positional coincidence with the expected coordinates.

\begin{figure}
\figurenum{2} 
\label{maglp94420}
\plotone{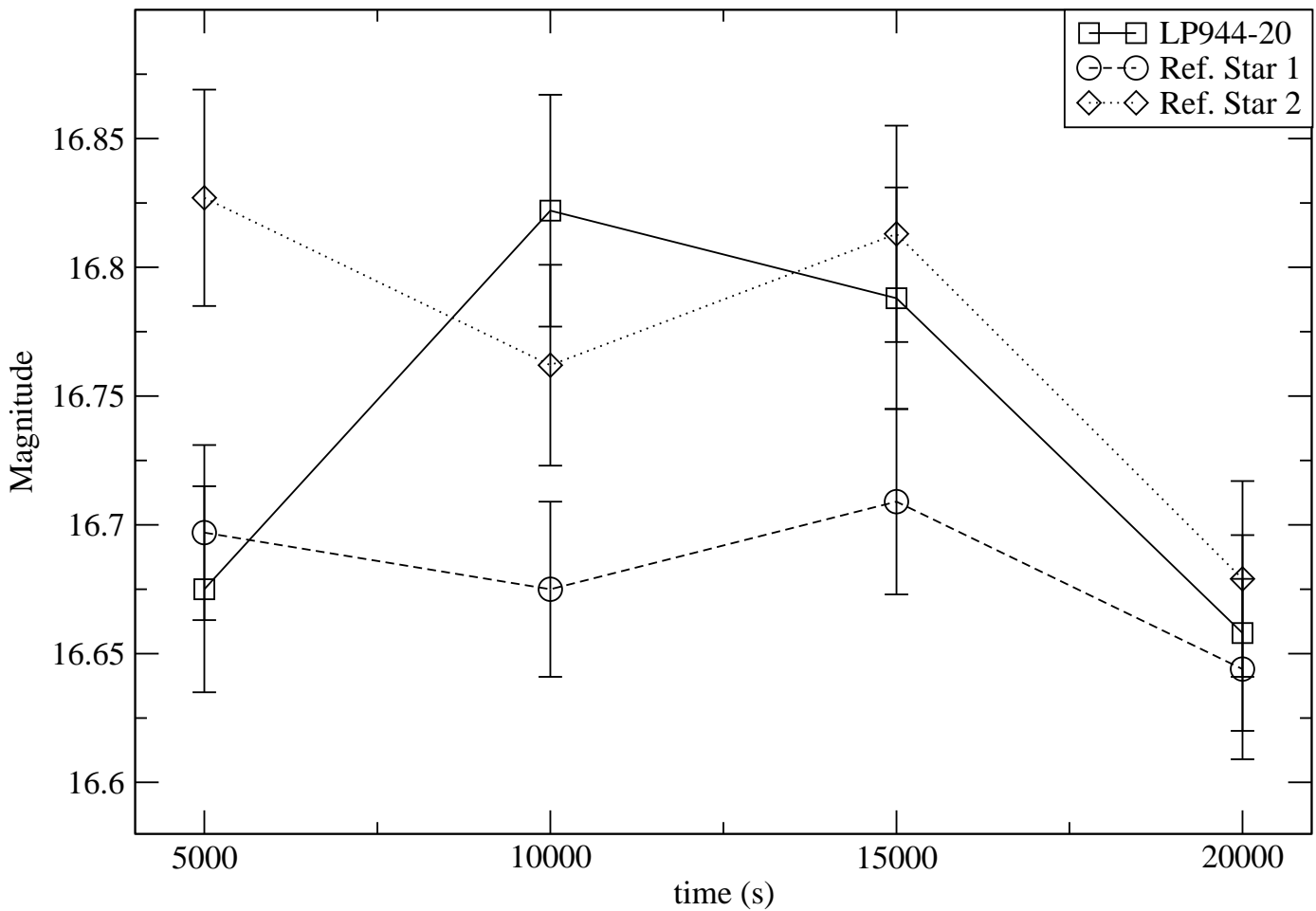}
\caption{Optical magnitude of \lp\ and of 2 reference stars in the field of view of the Optical Monitor.}
\end{figure}

We also used the four exposures given by the OM to look for optical variability. The magnitudes of the source identified as \lp\ and of two other stars in the field of view were calculated using the \emph{IRAF} \verb+phot+ task. This allowed us to plot the evolution of the magnitude during the whole observation. The result is shown in Figure \ref{maglp94420}.
The behavior of \lp's magnitude is different than that of the two other stars. 
We notice that whereas the magnitudes of the two reference stars are roughly correlated, \lp's magnitudes are not. 
We measure a mean magnitude of $V=$16.736 and a standard deviation of 0.081 for \lp. For reference star 1 we measure 
a mean magnitude of $V=$16.681 and  a standard deviation of 0.028. For reference star 2 we measure 
a mean magnitude of $V=$16.770 and  a standard deviation of 0.067. We conclude that \lp\ may be variable in the OM data. The variability of \lp, however, is only slightly higher 
than that of reference star 2. We cannot make a definitive claim that \lp\ is variable in our data because of the lack of enough reference stars in the field of view.

\subsection{X-ray Data and Analysis}

The European Photon Imaging Camera (\emph{EPIC}) MOS1 \&  MOS2 were used. 
The total ontarget exposure time in each camera was 48724 s in prime full window mode with 2.5 s time resolution. 
The EPIC PN was used in prime full window mode with 73.4 ms time resolution and ontarget exposure 46618~s. 
Each EPIC instrument was used with the thin filter, which suppresses the optical contamination up to $m_{V}$ of 18 (MOS) 
or $m_{V}$ of 17 (PN).
Both MOS \& PN detectors have a circular field of view of $30'$ diameter. For more details on \xmm\ and its instruments, we refer to \citet{xmmuhb}.

\begin{figure}
\figurenum{3}
\label{mospn1}
\plotone{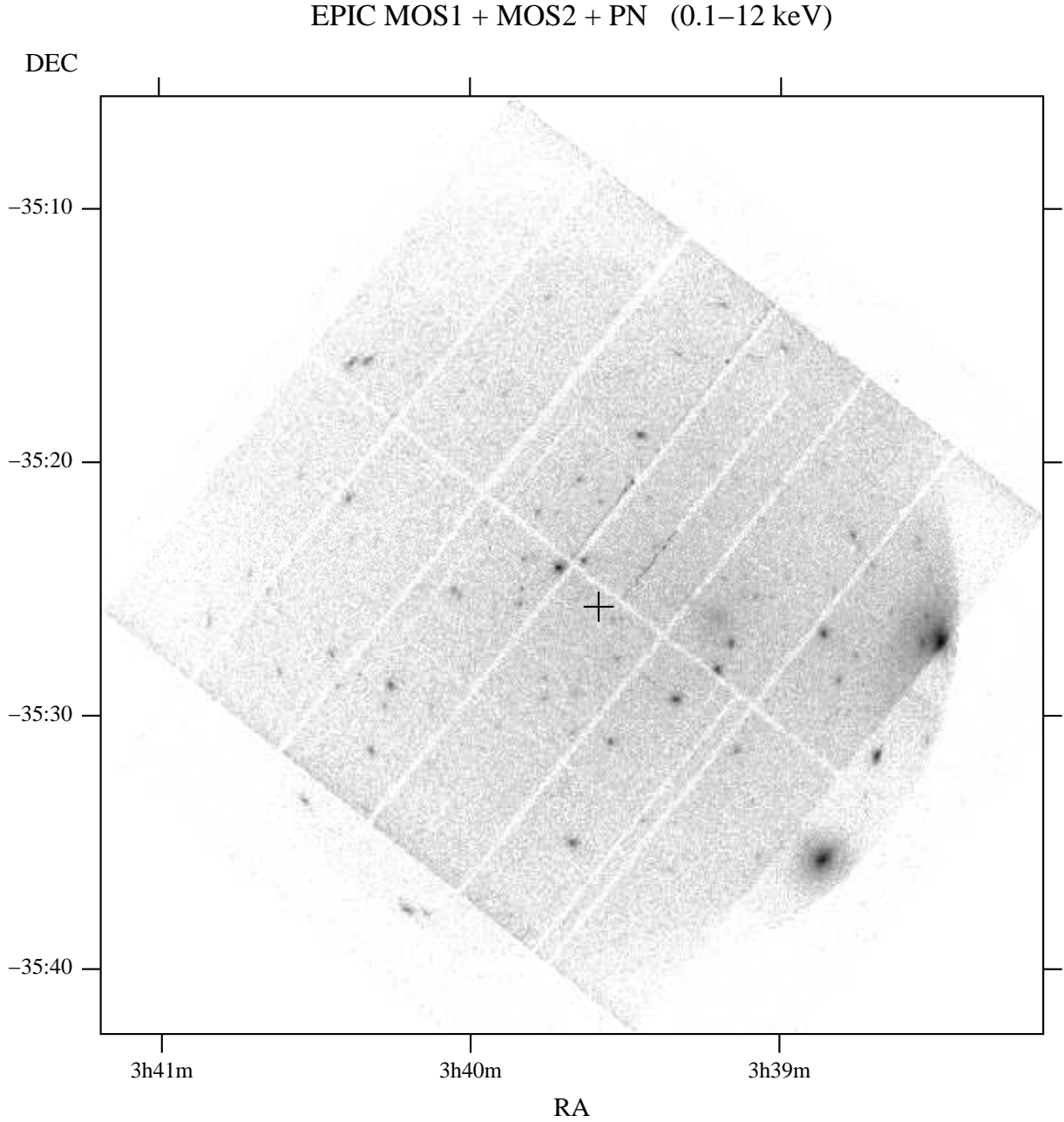}
\caption{\small Combined image from all EPICs MOS \& PN instruments in the energy range ($0.1-12.0$ $keV$). 
The figure shows the entire field of view. The square chips of the EPIC PN and the circular field of view of MOS1 \& 2 appear on this figure. A plus indicate the expected position of \lp.}
\end{figure}

The first step was to search for \lp\ in the field of view of each EPIC MOS \& PN detector. 
We plotted the coordinates of \lp\ on each image obtained by each detector, but we did not detect any source.  
To be sure that we were using the most sensitive exposure, we used the image given by the pipeline which combines data from all 
EPIC's MOS \& PN in the entire energy range ($0.1-12.0$ $keV$), as shown in Figure \ref{mospn1}.

On these figures, the X-ray source closest to the expected position is offset by $\delta (R.A.)=2.5'' \pm 0.1''$ and $\delta (Dec)=-28.37'' \pm 0.1''$. This offset is large in comparison with the $0.4'' \pm 0.1''$ offset observed with the optical data. 
So at this point we already knew that \lp\ was not detected.

We performed differential astrometry between the \xmm\ and \emph{Chandra} images. 
This allowed us to identify all the X-ray sources in our field of view in an area of $20'' \times 10''$ centered on 
the expected coordinates of \lp. We were then able to conclude that \lp\ was not detected during the \xmm\ observation.  

Since the target was not detected, the first conclusion is that there was no significant X-ray flare during this $48$ ks observation. 
Nevertheless, it is interesting to calculate the upper limit on the time average quiescent X-ray luminosity.
We did the calculation of the upper limit of X-ray flux with each of the 3 images of the preprocessing products, corresponding to the 3 exposures obtained with the EPIC MOS1, MOS2 \& PN detectors on the entire energy range ($0.1-12.0$ keV).
To calculate a $3 \sigma$ flux upper  limit, we considered an area of twice the FWHM ($=2$ pixels) around the expected position of 
\lp. In this area we calculated the mean value of $counts \cdot pixel^{-1}$ and the corresponding standard deviation $\sigma$ using the IRAF\footnote{\emph{IRAF} is distributed by National Optical Astronomy Observatory, which is operated by the Association of Universities for Research in Astronomy, Inc., under contract with the National Science Foundation.} \verb+phot+ task, which gives count rates corrected for background. The results are listed in Table \ref{table1}. As the EPIC PN camera is the most sensitive detector, 
we therefore only consider its result to calculate the flux upper limit. 

To convert this count rate to flux, we followed \citet{xmmuhb} instructions, by using the \pimms\footnotemark[3] software. As \lp\ was not detected during the observation, we were not able to fit its spectrum in order to know its spectral type. We adopted the results of \citet{rutledge2000} summarized in Table \ref{table2}. The HI column density was calculated using the \emph{nH}\footnotemark[3] software: $n_{H}=1.31 \times 10^{20}$$cm^{-2}$. 

\footnotetext[3]{\emph{nH} and \emph{PIMMS} (Portable, Interactive, Multi-Mission Simulator) are distributed by the NASA's HEARSAC}

The results for the upper limit of flux and X-ray luminosity (3$\sigma$, assuming Gaussian statistics) are listed in Table \ref{table3}. The uncertainties on these values are Poisson (counting statistics) plus $\sim 10 \%$ spectral uncertainty, according to \citet{rutledge2000}. As their best fit on the spectrum was obtained with the Raymond--Smith thermal plasma model, we finally kept the corresponding 3 $\sigma$ flux as the upper limit on the time average quiescent X-ray luminosity for \lp: $L_{X}<3.1 \times 10^{23} ergs \cdot s^{-1}$, equivalent to a luminosity ratio upper limit of $\log{(L_{X}/L_{bol})} \le -6.28$.  

\section{Discussion}

The first observation of \lp\ in the X-ray regime by \citet{neuhauser99} with the \emph{ROSAT} satellite placed an upper limit of $\log{(L_{X}/L_{bol})} \le -4.17$ after a 220 ks \emph{ROSAT---HRI} observation. The second X-ray observation \citep{rutledge2000} placed a better upper limit of $\log{(L_{X}/L_{bol})} \le -5.70$ after the 44 ks \emph{Chandra---ACIS-S} observation before the flare. Our new value improves this upper limit of quiescent X-ray emission at $\log{(L_{X}/L_{bol})} \le -6.3$. The corresponding upper limit of X-ray luminosity ($L_{X} \le 3.1 \times 10^{23} ergs \cdot s^{-1}$) is thus improved by a factor 3. This value is close to the solar X-ray emission level  
during its maxima of activity.

\begin{figure}
\figurenum{4}
\label{logL_vs_sptype}
\plotone{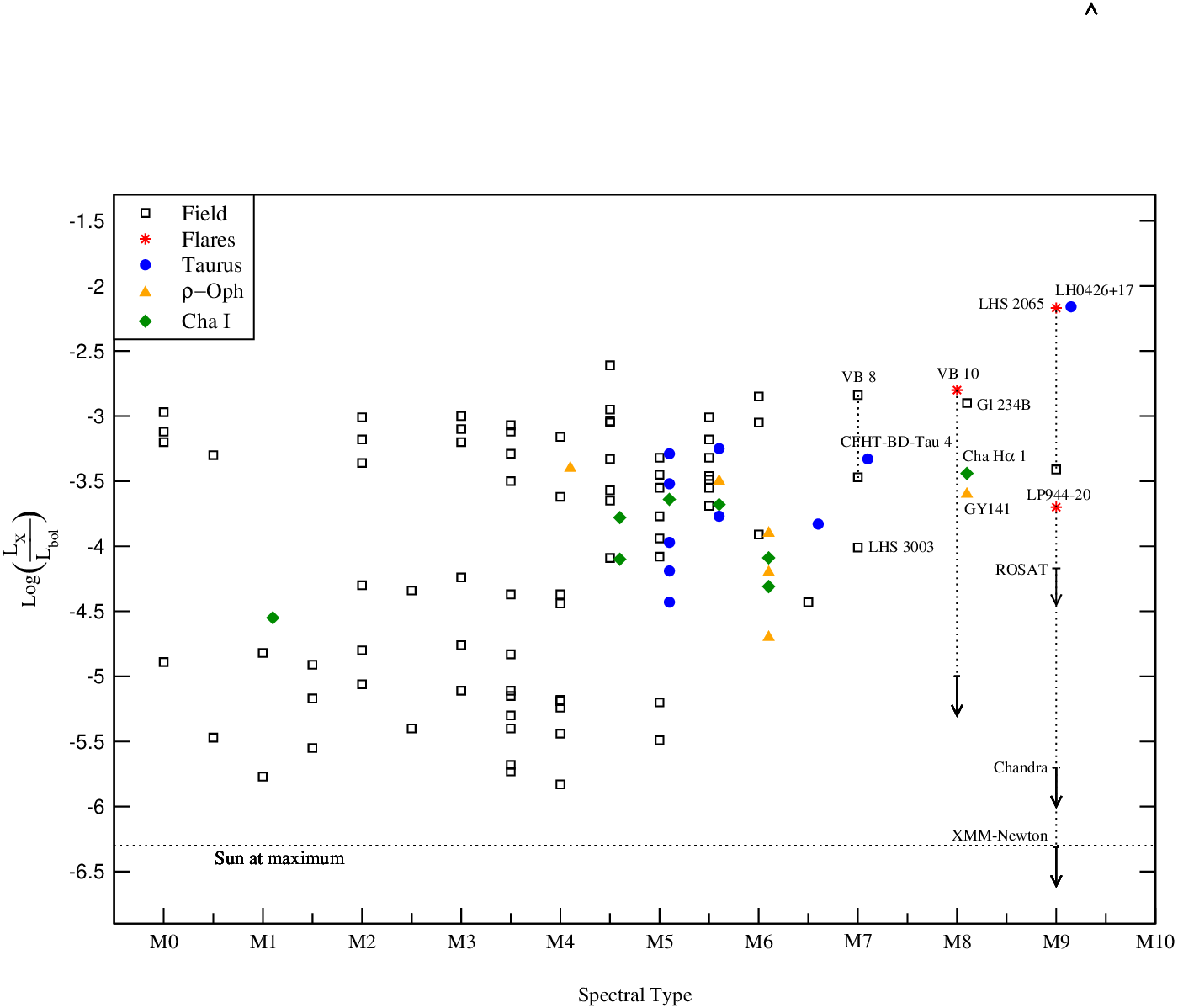}
\caption{Distribution of X-ray luminosity of dM dwarfs as a function of spectral type. 
The squares represent values obtained for field dwarfs (\citet{fleming1993, fleming1995, fleming2000}; \citet{giampapa1996}; \citet{rutledge2000}). 
The circles, diamonds and triangles denote young objects in star-forming regions.  
The stars represent the flaring objects at their maximum with the corresponding value or upper limit on the quiescent emission linked by a dotted line. To avoid confusion, the Taurus object's spectral types have been shifted by 0.1 subclass.}
\end{figure}

In Figure \ref{logL_vs_sptype} we display the dependence of X-ray activity versus spectral type for M dwarfs. 
The upper envelope of X-ray activity is remarkably independent of M subclass.  
There is no obvious connection between age and X-rays. The M dwarfs in the ChaI, $\rho$Oph and 
Taurus star-forming regions show X-ray emission 
comparable to much older field dwarfs \citep{neuhauser99, mokler2002}. We conclude that there is no evidence 
for a decline in coronal emission with increasing age in fully convective M dwarfs.  

We have estimated the rate of X-ray flare occurrence in \lp\ by adding 287.5 ks of 
\emph{ROSAT} data analyzed by \citet{neuhauser99}, 43.8 ks of \emph{Chandra} data analyzed by 
\citet{rutledge2000} and 51.7 ks of \xmm\ data (this paper). The duration of the only flare 
so far observed was 2.76 ks \citet{rutledge2000}. Thus, we get an X-ray flare frequency of 0.72\% for flares with luminosity Log$(L_X/L_{bol})>-$2.5. 
Weaker flares could not have been detected by \emph{ROSAT}. Thus, we derive the 
duty cycle of flares with luminosity comparable with the flare detected by 
\emph{Chandra} [Log$(L_X/L_{bol})=-$4.1] to be about 6\%.

The published VLA radio data have a duration of 79.9 ks \citep{berger2001}. 
Three flares were detected with a total duration of $\sim$10.4 ks. 
Hence, we estimate a radio flare frequency of about 13\%. 
It is not known to what X-ray luminosity the radio flares correspond, but we can 
set an upper limit using the \xmm\ data. If we assume a flare duty cycle of 
13\%, the flare X-ray luminosity has to be Log$(L_X/L_{bol})<-$5.39 to be consistent 
with our non detection.   

\lp\ is a relatively inactive object. \citet{tinney98} reported H$_\alpha$ weakly in emission. 
\citet{berger2001} estimated a surface magnetic field strength of 
B$\sim$5G assuming synchrotron emission during the radio flares. 
For comparison, active dM stars have magnetic field strengths of a few kG, 
solar flares have B$\sim$100G, and Jupiter has an average B$\sim$10G \citep{hide76}. 
Our low upper limit on the quiescent X-ray emission of \lp\ is consistent with a weak magnetic field. 
The detection of low-amplitude photometric variability in \lp, and the low-level of X-ray emission in this object, 
suggests the presence of surface thermal inhomogeneities due to nonmagnetic clouds, as discussed by \citet{tinney99} 
for \lp\ and by \citet{martin2001} for the M9.5 dwarf BRI0021-0214. 
Recent theoretical works  \citep{gelino02,mohanty02} show  
that in ultracool dwarfs there should not be much interaction between the gas and the magnetic field because 
the Reynolds number is very low throughout the photosphere. 

\lp\ appears to be an object with substellar mass, weak magnetic field, fast rotation and cloudy atmosphere. We speculate 
that the radio activity may be enhanced by the presence of a close planetary-mass companion (Io-type), which provides  
a continuous supply of ionized particles to the ionosphere of \lp. 
Radial velocity variability has been detected by \citet{guenther03} which could 
be due to a planetary-mass close companion, but they did not have enough data to 
derive an orbital period.

\newpage


\acknowledgments

This research has made use of the Simbad \footnote{\url{http://simbad.u-strasbg.fr/}} database, operated at CDS, Strasbourg, France. Support for this project has been provided by NASA grant NAG5-9992. We thank our collaborators Gibor Basri, Lars Bildsten and Bob Rutledge, and an anonymous referee for helpful discussions.


\clearpage


\begin{deluxetable}{l c c c}
\tablecaption{Count rates \label{table1}}
\tablewidth{0pt}
\tablehead{
\colhead{Instrument} & \colhead{Counts/pixel (average)}  & \colhead{Standart deviation $\sigma$ (counts/pixel)} & \colhead{ $3 \sigma$ (counts/pixel/s)}
}
\startdata
EPIC PN & $4.53$ & $2.00 $ & $1.3 \times 10^{-4}$  \\
EPIC MOS1 & $1.59$ & $1.18$ & $7.3 \times 10^{-5}$ \\
EPIC MOS2 & $1.73$ & $1.24$ & $7.6 \times  10^{-5}$ \\

\enddata
\end{deluxetable}

\begin{deluxetable}{l l}
\tablecaption{Specral parameters of \lp\ \label{table2}}
\tablewidth{0pt}
\tablehead{
\colhead{Model} & \colhead{Associated Feature}
}
\startdata
Black Body & $k_{B} \times T = 0.17$ keV \\
Power Law & $\alpha_{photon}=2.6$ \\
Thermal Plasma & $k_{B} \times T = 0.26$ $keV$ 
\tablerefs{ \footnotesize \citet{rutledge2000}}
\enddata 
\end{deluxetable}

\begin{deluxetable}{l c c c}
\tablecaption{X-ray luminosity upper limit \label{table3}}
\tablewidth{0pt}
\tablehead{
\colhead{Model} & \colhead{Flux (\scriptsize $10^{-17} erg.cm^{-2}.s^{-1}$ \normalsize)} & \colhead{$\log{L_{X}}$ (\scriptsize$ergs.s^{-1}$  \normalsize)} & \colhead{$\log{(\frac{L_{X}}{L_{bol}})}$}
}
\startdata
Black Body & $\le 1.1$ & $ \le 23.50$ & $\le -6.27$\\
Power Law & $\le 1.3$ & $\le 23.59$ & $\le -6.19 $ \\
Thermal Plasma & $\le 1.0$ & $\le 23.49$ & $\le -6.28$
\tablerefs{ \footnotesize \citet{tinney96} for $L_{bol} \sim 6 \times 10^{29}ergs.s^{-1} $}
\enddata 
\end{deluxetable}

\end{document}